\documentstyle[tighten,aps,psfig]{revtex}

\twocolumn

\newcommand{\be}{\begin{equation}}
\newcommand{\ee}{\end{equation}}
\newcommand{\ba}{\begin{eqnarray}}
\newcommand{\ea}{\end{eqnarray}}

\newcommand{\ep}{\epsilon}

\begin{document}

\title{
%
%
\[ \vspace{-2cm} \]
\noindent\hfill\hbox{\rm  } \vskip 1pt
\noindent\hfill\hbox{\rm Alberta Thy 12-00} \vskip 1pt
\noindent\hfill\hbox{\rm SLAC-PUB-8730} \vskip 1pt
\noindent\hfill\hbox{\rm hep-ph/0012053} \vskip 10pt
%
%
Expansion of bound-state energies in powers of $m/M$
}

\author{Andrzej Czarnecki\thanks{
e-mail:  czar@phys.ualberta.ca}}
\address{
Department of Physics, University of Alberta\\
Edmonton, AB\ \  T6G 2J1, Canada}

\author{Kirill Melnikov\thanks{
e-mail:  melnikov@slac.stanford.edu}}
\address{Stanford Linear Accelerator Center\\
Stanford University, Stanford, CA 94309}
\maketitle

\begin{abstract}
We describe a new approach to computing energy levels of a
non-relativistic bound-state of two constituents with masses $M$ and
$m$, by a systematic expansion in powers of $m/M$.  After discussing
the method, we demonstrate its potential with an example of the
radiative recoil corrections to the Lamb shift and hyperfine splitting
relevant for the hydrogen, muonic hydrogen, and muonium.  A
discrepancy between two previous calculations of ${\cal
O}(\alpha(Z\alpha)^5 m^2/M)$ radiative recoil corrections to the Lamb
shift is resolved and several 
new terms of ${\cal O}(\alpha(Z\alpha)^5
m^4/M^3)$ and higher  are obtained.
\end{abstract}


The theory of non-relativistic bound states in QED remains an
important source of information about fundamental physical parameters,
like the fine structure constant and the masses of the electron, muon
and proton, among many others \cite{Mohr99}. Simple atoms, which are
being studied in laboratories, differ significantly in the ratios of
their constituent masses.  Two situations can be
distinguished.  The first one is the case when the masses of the two
constituents of the bound state are equal, with the positronium as the
most important example.  The second case is a bound state with two
very different masses, e.g. hydrogen, muonium, muonic hydrogen.  Both
situations represent two special limits of a general mass ratio case.
In both limits certain simplifications are possible. In the context of
this Letter, the case of equal constituent masses was discussed to
some extent in \cite{Czarnecki:1998zv,Czarnecki:1999mw}.  In this
Letter we consider the case when the masses of the constituents differ
significantly from one another.

Our main goal is a practical algorithm which allows a calculation of
the bound-state energy levels in a given order of perturbation theory
(in $\alpha$ and $Z\alpha$) as an expansion in powers and logarithms
of $m/M$ with an arbitrary precision.  The opposite situation,
i.e. calculation of the energy levels to all orders in $\alpha$ but in
a fixed order in the ratio $m/M$, has been studied in the literature
\cite{braun,shabaev,Yelkhovsky:1994ag,PG95}.

In many practical situations only the first few terms of the expansion
in $m/M$ are required.  Nevertheless, we believe that it is useful to
construct such an algorithm in its generality. First, higher
corrections in the ratio $m/M$ might become relevant.  For example, in
the muonic hydrogen $m/M$ corresponds to $m_\mu/M_p\simeq 0.113$, not
a very small parameter.  In exotic hadronic atoms, such as pionic
hydrogen, that ratio might be even larger.  In hydrogen and muonium,
where $m/M$ is smaller, the very high precision of experiments
warrants a precise computation of the recoil effects.  Second, a
complete algorithm means that one can obtain the whole series in $m/M$
at once and additional cross checks become possible.

We first recall that non-relativistic bound-state energies can be
computed using an effective field theory \cite{Caswell:1986ui}. In
\cite{Czarnecki:1998zv,Czarnecki:1999mw} we have shown how dimensional
regularization facilitates this approach.  One has to distinguish two
different contributions. The first one is the contribution of the
relativistic region in the loop-momentum integrals; in what follows we
will refer to these contributions as ``hard'' contributions. This is
usually obtained as a Taylor expansion of the relevant scattering
amplitudes in spatial momentum components of external particles (which
can be taken on-shell).

Second, there is the so-called ``soft'' contribution, 
given by usual time-independent  perturbation
theory in quantum mechanics. An important point to note is that the
soft contribution can in general be easily evaluated for arbitrary
masses of constituents. As one can see from the Schr\"odinger 
equation, this is so  because the essential
dynamics of a non-relativistic bound state 
is governed by the reduced mass of the system rather 
than by the masses of individual constituents.  In this respect, 
for the soft contributions the relation between the two masses 
is not very important and once the equal mass 
case has been solved, the rest follows easily.

Therefore, in the situation where the two masses are different, the
real problem is in computing the hard contribution and this is what we
are going to discuss in this Letter.  We will show that there is a
simple way to expand the hard scattering diagrams in powers of
$m/M$. The essential advantage of this method is that it can be
automated and many terms of the expansion can be easily computed.  The
only limitation is the available computer power; as a matter of
principle, infinitely many terms in the $m/M$ expansion can be
obtained.  High-performance symbolic algebra software is of great
help in such computations (we use FORM \cite{form3}).

The remainder of this Letter is organized as follows. First
the method is described in detail. Next, we compute
the $\alpha(Z\alpha)^5$ radiative recoil corrections to both the 
Lamb shift and the hyperfine splitting up to the 
fourth order in the expansion in $m/M$. Finally, we present our
conclusions.

The method we are going to discuss is based entirely on using 
dimensional regularization. Note that for consistency 
one also needs the soft contribution in dimensional 
regularization; as we mentioned earlier this 
part of the problem is well understood
\cite{Czarnecki:1998zv,Czarnecki:1999mw}.

The hard diagrams should be evaluated exactly at the threshold (zero
relative velocity of the constituents).  For this reason, the relevant
Feynman integrals depend on only two scales, $m$ and $M$. Since we are
interested in their expansion in $m/M$, it is useful to be able
to expand the {\it integrands}, so that only homogeneous one-scale
integrals have to be evaluated.  Once this is achieved, the
calculations simplify dramatically.

\begin{figure}[htb]
\begin{tabular}{cc}
\psfig{figure=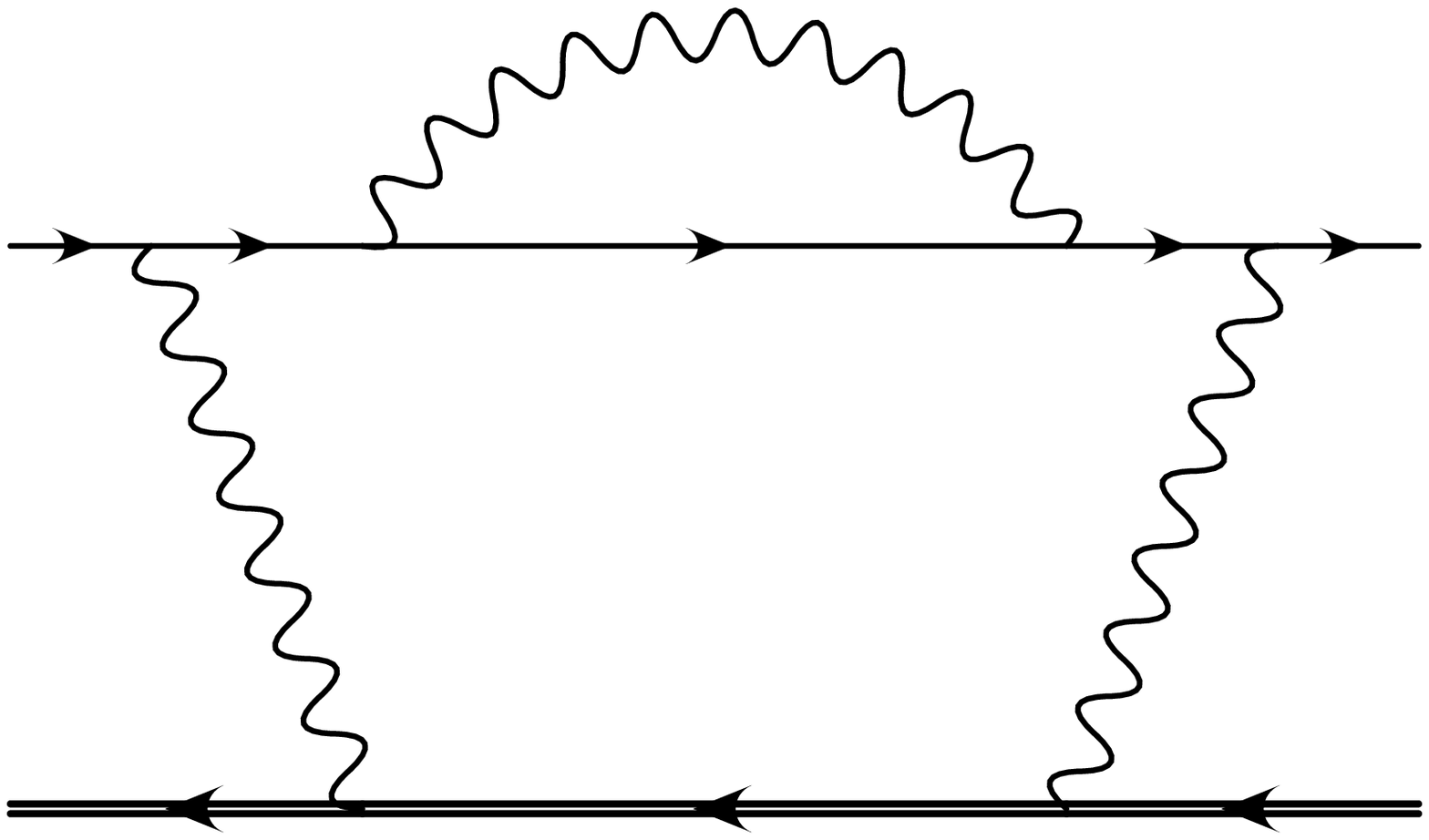,width=35mm} &
\hspace*{5mm}
\psfig{figure=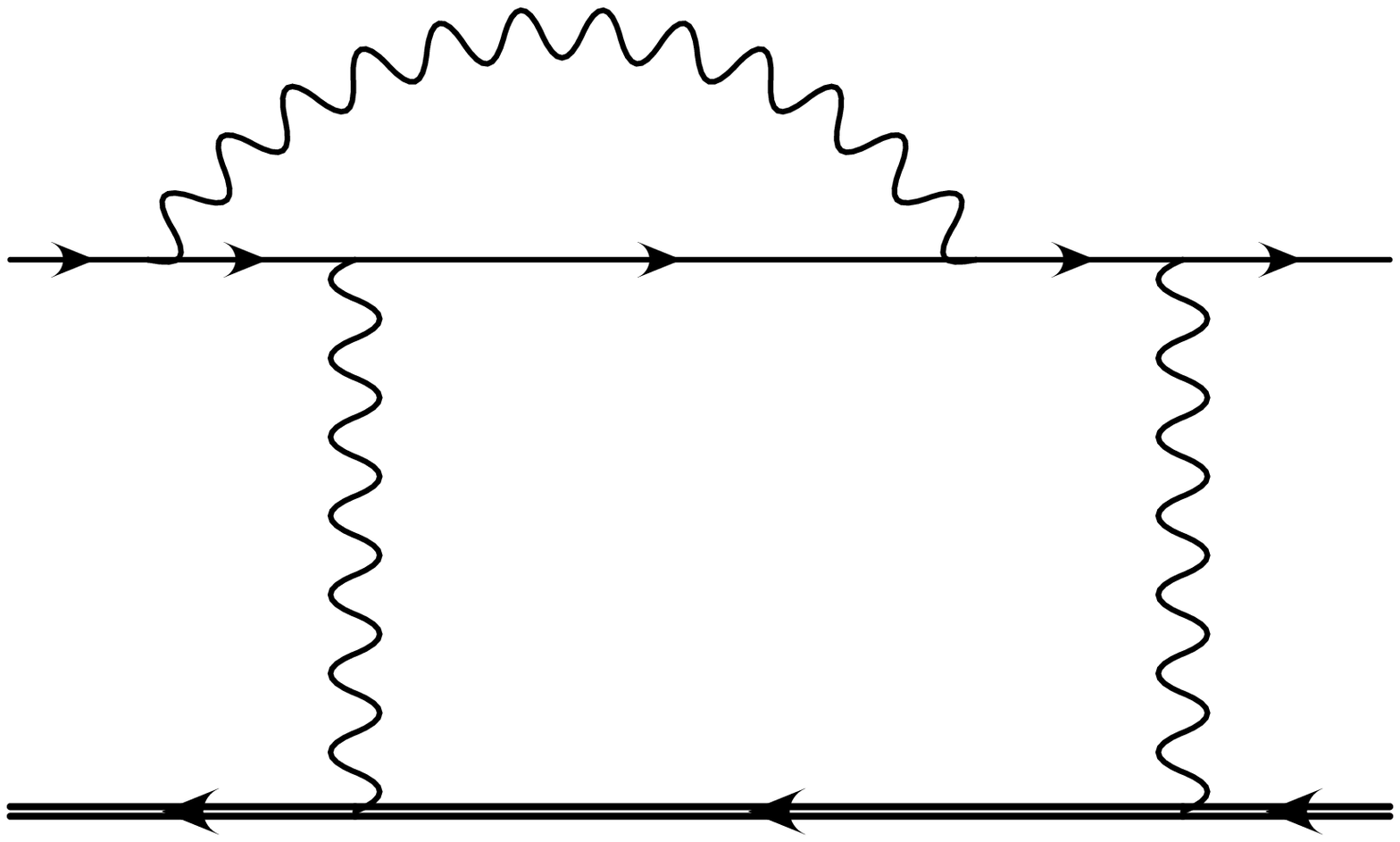,width=35mm} 
\end{tabular}

\vspace*{5mm}

\hspace*{19mm}
\psfig{figure=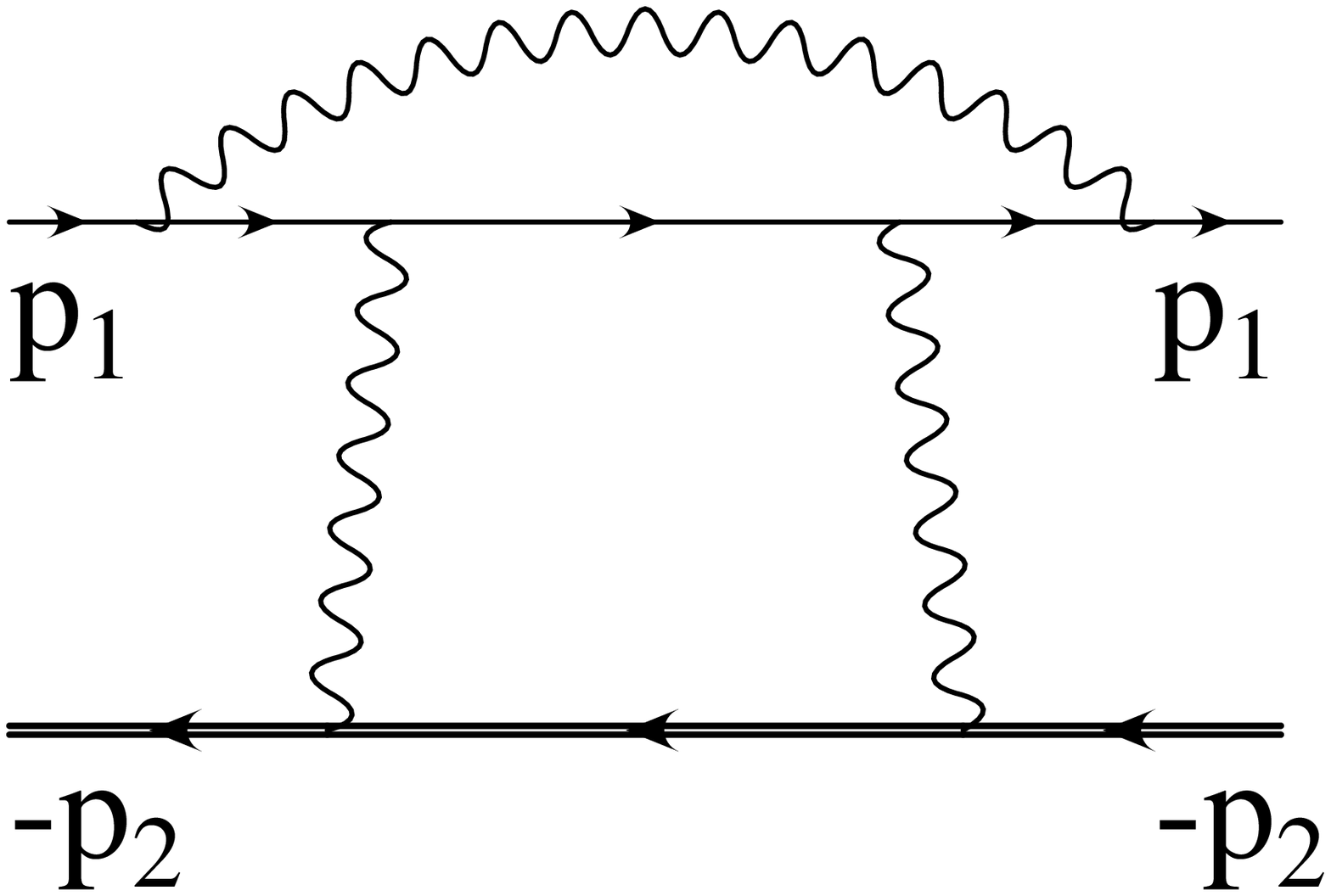,width=43mm}

\vspace*{3mm}

\caption{The forward-scattering radiative-recoil diagrams. The bold
line represents the heavy constituent of the bound-state (e.g. proton
if we consider hydrogen) and the thin line --- the light one (an
electron). Diagrams with the crossed photons in the $t$-channel are
not displayed.}
\label{fig1}
\end{figure}

Our method is motivated by the known procedure which permits an
expansion of 
Feynman diagrams in large momenta and masses
\cite{Chetyrkin91,Tkachev:1994gz,Smirnov:1995tg}.  Although 
originally formulated in a different way, that procedure can be
reformulated more practically using the notion of momentum regions.
To arrive at the result one should follow a sequence of steps 
\cite{beneke}: $(1)$
determine large and small scales in the problem; $(2)$ divide the
entire integration volume into regions where each loop momentum is of
the order of some of the characteristic scales; $(3)$ in every region
perform a Taylor expansion in the parameters which are small in the
given region; $(4)$ after the expansion, ignore all the constraints on
the regions and perform the integration over the entire integration
volume; $(5)$ add the contributions of different regions to obtain the
final result.  The only step in this sequence which might appear
counter-intuitive is the step $4$, since one may suspect some double
counting. The reason why that does not happen is that the scale-less
integrals vanish in the dimensional regularization.  This in turn
implies that the results obtained from the integrals over different
regions are different analytic functions of the parameters of the
problem.  Below this procedure will be demonstrated in some detail.

To illustrate the method we focus on the last diagram
in Fig.~\ref{fig1} and consider the following scalar
integral:
\ba
&&\int
\frac {[{\rm d}^D k_1] [{\rm d}^D k_2]}
{{(k_1^2)}^2 (k_2^2) {(k_2^2 + 2p_1 k_2 + i \delta )}^2}
\nonumber \\
&& \times \frac {1}{
\left[(k_1+k_2)^2 + 2p_1(k_1+k_2) + i \delta\right]
(k_1^2 - 2p_2k_1 + i \delta)}.
\label{exint}
\ea
Here $[{\rm d}^D k]$ stands for ${\rm d}^D k/(2\pi)^D$, $p_1 \equiv mQ$,
$p_2 \equiv MQ$, where $Q=(1,0,0,0)$ is the time-like unit
vector. Only the
relevant infinitesimal imaginary parts of the propagators have been
displayed.  We are going to illustrate the 
expansion of the integral in Eq.~(\ref{exint}) in powers of $m/M$ following 
the five steps outlined above.

There are four momentum regions to be considered. In the first one
all the momenta are of the order of the large mass $M$. In this
case one can expand the electron propagators in $mQk_i$. The resulting
integrals are all of the form:
\be
\int \frac {[{\rm d}^D k_1] [{\rm d}^D k_2]}
{{(k_1^2)}^{a_1} {(k_2^2)}^{a_2} {(k_1+ k_2)}^{2a_3}
{(k_1^2 - 2p_2k_1)}^{a_4}},
\ee
with some integer powers $a_i$. One immediately recognizes that
all these integrals are identical with the general two-loop self-energy
integrals of the particle with the mass $M$ for which the general
solution is known \cite{bgs91}.

Next, there are two momentum regions where either $k_1 \sim M$ and
$k_2 \sim m$ or $k_1 \sim M$ and  $k_2 \sim M$, but $k_1 + k_2 \sim m$.
It is then easy to see that after a Taylor expansion in the small
variables, the integral  factorizes into a product 
of two simple one-loop integrals.

The fourth region is determined by the condition
$k_1 \sim  k_2 \sim m$. In this case the heavy particle propagator 
can be
expanded in powers of $k_1^2$  and in essence it becomes a {\it static} 
propagator. The  general integral in this case has the form
\ba
&& J =
\int
\frac {[{\rm d}^D k_1] [{\rm d}^D k_2]}
{{(k_1^2)}^{a_1} {(k_2^2)}^{a_2} {(k_2^2 + 2p_1 k_2 + i \delta )}^{a_3}}
\nonumber \\
&& \times \frac {1}{
{\left[(k_1+k_2)^2 + 2p_1(k_1+k_2) + i \delta\right]}^{a_4}
{(2p_2k_1 - i \delta)}^{a_5}}.
\label{ss}
\ea
Such integrals represent the only new type required
for this calculation and the easiest way to solve them is to
employ the integration-by-parts techniques \cite{Tkachev:1981wb,che81}. 
Any integral $J$ can be algebraically
expressed as a combination of the  
two-loop on-shell self-energy integrals and 
four new master integrals. The latter
are the only integrals we have to compute, but this can be easily
accomplished with help of  Feynman
parameters.  The results read
\ba
&& J_{1}^{\pm} = \int \frac {[{\rm d}^D k_1] [{\rm d}^D k_2]}
{(k_1 Q -1 \pm i \delta) (k_2^2+i\delta) 
\left[ (k_1+k_2)^2 -1 + i \delta\right] }
\nonumber \\
=&& {1\over (4\pi)^D} \left[
2\Gamma(1-\ep) \Gamma(3\ep-2) B(4\ep-3,2\ep-1) 
\right. \nonumber \\
&& \left. 
-(1 \mp 1)
 \sqrt{\pi}
\Gamma\left(2\ep-{3\over 2}\right) 
B\left(\frac {5}{2}-3\ep,-\frac {1}{2}+\ep\right)
 \right],
\ea
\ba
J_{2}^{\pm} =&& \int \frac {[{\rm d}^D k_1] [{\rm d}^D k_2]}
{(k_1 Q \pm i \delta) (k_2^2 - 1 +i\delta) 
\left[ (k_1+k_2)^2 -1 + i \delta\right] }
\nonumber \\
&&=
\pm \frac {\sqrt{\pi}}{(4\pi)^D}
\Gamma\left(2\ep-\frac {3}{2}\right) 
B\left(\frac {5}{2}-3\ep,-\frac {1}{2}+\ep\right).
\ea
Let us add that all momentum regions other than the ones we
discussed lead to scale-less integrals and are therefore not relevant. 
This concludes the construction of our expansion algorithm.

We have applied this algorithm to compute the ${\cal O}(\alpha (Z\alpha)^5)$
radiative recoil corrections to the Lamb shift and the hyperfine
splitting of a general QED bound state composed of two spin-1/2 particles 
with the masses  $m$ and $M$.  It is well known that in this case 
the soft contribution is absent and the hard corrections 
shown in Fig.~1 are the only diagrams we have to consider.
We have done the calculation in a general covariant gauge; the 
cancellation of the gauge parameter dependence serves as 
a check of the computation.

For the $S$-wave ground state energy $E$ we define:
\be
E = E_{\rm aver} + \left ( \frac {1}{4} - \delta_{J0} \right ) E_{\rm hfs},
\ee
where $J=0,1$ is the total spin of the two fermions forming the bound state.

For the hyperfine splitting we obtain:
\ba
&& \delta E_{\rm hfs}^{\rm rad~rec} \simeq \frac {8 (Z\alpha)^4 \mu^3}{3mM} 
\alpha (Z\alpha) 
\left \{ \ln 2 - \frac {13}{4}
\right.
\nonumber \\
&& \left.
+ \frac {m}{M} \left ( \frac {15}{4\pi^2} \ln \frac {M}{m}
  + \frac {1}{2}
 + \frac {6 \zeta_3}{\pi^2} + \frac {17}{8 \pi^2} + 3 \ln 2 \right )
\right. \nonumber \\
&& 
 -\left({m\over M}\right)^2 \left ( \frac {3}{2} + 6 \ln 2 \right )
\nonumber \\ &&
+ \left({m\over M}\right)^3 \left (
\frac {61}{12 \pi^2 } \ln^2 \frac {M}{m}
+ \frac {1037}{72 \pi^2 } \ln \frac {M}{m}
\right. 
\nonumber \\
&&\qquad  \qquad  
\left.+  \frac {133}{72}
+ \frac {9\zeta_3}{2\pi^2} + \frac {5521}{288\pi^2} + 3 \ln 2
\right )
\nonumber \\
&&
 - \left({m\over M}\right)^4 \left(   {163\over 48} + 6\ln 2 \right)
\nonumber \\
&&
   + \left({m\over M}\right)^5 \left( 
{331\over 40\pi^2}\ln^2{M\over m}
 + {5761\over 300\pi^2}\ln{M\over m}  +{691\over 240}  
\right.  
\nonumber \\
&&\left.\left. \qquad \qquad
+ {9\zeta_3 \over 2\pi^2} + {206653\over 8000\pi^2} + 3\ln 2
\right)
 \right \}, 
\ea
where $\mu = mM/(m+M)$ is the reduced mass of the bound state.

For the spin-independent energy shift  we find
\ba
&& \delta E_{\rm aver}^{\rm rad~rec} \simeq \alpha (Z\alpha)^5
 \frac {\mu^3}{m^2} \left \{
  \frac {139}{32} - 2 \ln 2
\right. \nonumber \\
&& \left.
+ \frac {m}{M} \left ( \frac {3}{4} + \frac {6 \zeta_3}{\pi^2}
    - \frac {14}{\pi^2} - 2 \ln 2  \right )
\right. \label{lamb} \nonumber \\
&& 
+ \left({m\over M}\right)^2 \left (  - \frac {127}{32} + 8\ln 2 \right )
\nonumber \\ &&
+\left({m\over M}\right)^3
 \left (  - \frac {8}{3 \pi^2 } \ln^2 \frac {M}{m}
 - \frac {55}{18 \pi^2 } \ln \frac {M}{m}+ \frac {47}{36}
\right. 
\nonumber \\
&& \left . 
\qquad \qquad  -  \frac {3\zeta_3}{\pi^2} - \frac {85}{9\pi^2 }
 - 2 \ln 2 \right )
\nonumber \\
&&
+
\left({m\over M}\right)^4 \left(  - {55\over 24} + 4\ln 2 \right)
\nonumber \\
&&
 + \left({m\over M}\right)^5 \left({37\over 60\pi^2}\ln^2{M\over m}  
+ {29\over 900\pi^2}\ln{M\over m} + {1027\over 360}
\right.
\nonumber \\ &&
\left. - {3\zeta_3\over \pi^2} - {370667\over 36000\pi^2} - 2\ln 2 \right)
\nonumber \\ &&
 \left.        + \left({m\over M}\right)^6 \left(  - {67\over 20} + 4\ln 2 \right)
 \right \}. 
\ea

To our knowledge the terms ${\cal O}(m^3/M^3)$ and higher are new for
both $E_{\rm hfs}$ and $E_{\rm aver}$, while the other terms have been
obtained previously.  In addition, the coefficient of the ${\cal O}(m/M)$
term in Eq.~(\ref{lamb}) has been subject of some controversy, since
two different numerical results have been reported,
\cite{Bhatt1,Bhatt2,Bhatt3} and \cite{KPLamb}.

Our result for this term,
\ba
&& \alpha (Z\alpha)^5
 \frac {\mu^3}{m^2} \frac {m}{M} \left ( \frac {3}{4}
+ \frac {6 \zeta_3}{\pi^2} - \frac {14}{\pi^2} - 2 \ln 2  \right )
\nonumber \\
&& \simeq -1.32402796~\alpha (Z\alpha)^5
 \frac {\mu^3}{m^2} \frac {m}{M},
\label{e10}
\ea
is in excellent agreement with the numerical result of
Ref.~\cite{KPLamb} where the coefficient 
$-1.324029(2)$ was obtained. 

The discrepancy in the ${\cal O}(m^2/M\alpha(Z\alpha)^5)$ corrections
to the Lamb shift reported in \cite{Bhatt1,Bhatt2,Bhatt3} and
\cite{KPLamb} has been the major source of the theoretical uncertainty
in the so-called isotope shift (apart from the uncertainty associated
with the proton and deuteron charge radii, see below), i.e.  the
difference between energies of $2S$ to $1S$ transitions in deuterium
and hydrogen: \be \Delta E = [E(2S)-E(1S)]_D - [E(2S)-E(1S)]_H.  \ee
Experimentally, $\Delta E$ is known with the uncertainty of about
$0.15~{\rm kHz}$ \cite{huber}; the theoretical uncertainty associated
with higher order QED effects and with the uncertainties in the
electron-to-proton and the electron-to-deuteron mass ratios is about
$1~{\rm kHz}$ each. On the other hand, the difference in the results
of Refs. \cite{Bhatt1,Bhatt2,Bhatt3} and \cite{KPLamb} leads to
$2.7~{\rm kHz}$ difference in $\Delta E$. Our result for this term,
Eq.~(\ref{e10}), removes this discrepancy in favor of the result of
Ref.~\cite{KPLamb}.

It is well known that the high accuracy of the experimental value of
$\Delta E$ cannot be used directly because of significant
uncertainties in the value of the proton and deuteron charge radii
which enter the theoretical formula for $\Delta E$. In this situation
the problem is usually turned around and one determines the difference
of the charge radii of the proton and deutron using $\Delta E$.  Here,
we do not pursue this topic any further.  The related phenomeno\-logy
can be extracted from Ref.~\cite{huber} (see also the recent review
\cite{eides}, where results of Ref.~\cite{KPLamb} should be used).

We have constructed an efficient algorithm which permits an expansion
of the energy levels of a bound state of two constituents with masses
$m$ and $M$ in powers of $m/M$.  This expansion is similar to,
although not identical with, the asymptotic expansions of Feynman
diagrams familiar from particle physics. We have demonstrated the
usefulness of this procedure by computing several  terms in the
$m/M$ expansion for the $\alpha (Z\alpha)^5$ radiative recoil
corrections to both the Lamb shift and the hyperfine splitting of a
general QED bound state.

Although we have only described a calculation of the radiative recoil
corrections, the method is clearly applicable to all other types
of corrections relevant for the bound states.  In particular, the pure 
recoil corrections can be treated in a similar way.  It remains
to work out the details in that case, but the principles
are clear.

One of the terms in our result for the radiative recoil corrections 
to the Lamb shift is the ${\cal O}(\alpha (Z\alpha)^5 \mu^3/(mM))$ 
term for which two different numerical results have been previously
reported. Our calculation  confirms the result of Ref.~\cite{KPLamb}.

Another aspect of this work might be related to higher number of
loops.  It is clear that the described method can be systematically
applied in  higher orders. Probably more important, it may
facilitate the extraction of terms enhanced by $\ln M/m$ which can be
determined from the singularities of the contributions of different
expansion regimes.  Since those singularities must cancel in the complete
result, their coefficients can be found by a partial calculation of
the divergent parts of those contributions which can be evaluated most
easily.

{\it Acknowledgments} 
This work was supported in part by the DOE under grant number
DE-AC03-76SF00515.


\end{document}